\documentclass[sigconf]{acmart}

\AtBeginDocument{%
  }

\setcopyright{acmlicensed}
\copyrightyear{2024}
\acmYear{2024}
\acmDOI{XXXXXXX.XXXXXXX}

\acmConference[RecSys '24]{18th ACM Conference on Recommender Systems}{October 14--18,
  2024}{Bari, Italy}

\acmBooktitle{18th ACM Conference on Recommender Systems (RecSys '24), October 14--18, 2024, Bari, Italy}

\copyrightyear{2024}
\acmYear{2024}
\setcopyright{rightsretained}
\acmConference[RecSys '24]{18th ACM Conference on Recommender Systems}{October 14--18, 2024}{Bari, Italy}
\acmBooktitle{18th ACM Conference on Recommender Systems (RecSys '24), October 14--18, 2024, Bari, Italy}\acmDOI{10.1145/3640457.3688051}
\acmISBN{979-8-4007-0505-2/24/10}

\begin{document}

\title{Sliding Window Training - Utilizing Historical Recommender Systems Data for Foundation Models}

\author{Swanand Joshi}
\email{swanandj@netflix.com}
\orcid{0009-0003-6316-4440}
\affiliation{%
  \institution{Netflix}
  \city{Los Gatos}
  \state{California}
  \country{USA}
}

\author{Yesu Feng}
\email{yfeng@netflix.com}
\orcid{0009-0008-4459-883X}
\affiliation{%
  \institution{Netflix}
  \city{Los Gatos}
  \state{California}
  \country{USA}
}

\author{Ko-Jen Hsiao}
\email{khsiao@netflix.com}
\orcid{0009-0003-6784-7556}
\affiliation{%
  \institution{Netflix}
  \city{Los Gatos}
  \state{California}
  \country{USA}
}

\author{Zhe Zhang}
\email{zhezhang@netflix.com}
\orcid{0000-0002-4056-1318}
\affiliation{%
  \institution{Netflix}
  \city{Los Gatos}
  \state{California}
  \country{USA}
}

\author{Sudarshan Lamkhede}
\email{slamkhede@netflix.com}
\orcid{0000-0001-8699-3776}
\affiliation{%
  \institution{Netflix}
  \city{Los Gatos}
  \state{California}
  \country{USA}
}

\begin{abstract}
  Long-lived recommender systems (RecSys) often encounter lengthy user-item interaction histories that span many years. To effectively learn long term user preferences, Large RecSys foundation models (FM) need to encode this information in pretraining. Usually, this is done by either generating a long enough sequence length to take all history sequences as input at the cost of large model input dimension or by dropping some parts of the user history to accommodate model size and latency requirements on the production serving side. In this paper, we introduce a sliding window training technique to incorporate long user history sequences during training time without increasing the model input dimension. We show the quantitative \& qualitative improvements this technique brings to the RecSys FM in learning user long term preferences. We additionally show that the average quality of items in the catalog learnt in pretraining also improves.
\end{abstract}


\received{3 June 2024}

\maketitle

\section{Motivation}
Foundation models are increasingly becoming popular in recommendation systems \cite{DBLP:journals/corr/abs-1904-06690}. The biggest leverage in using FMs for understanding what users are most interested in is user-item interaction sequences \cite{xia2023transact,huang2024foundation} . This pretrained FM is then used either to predict the next item to recommend in the recommendation system or to provide user and item representations for downstream use cases.

Oftentimes in industrial applications, foundation models that have inference time restrictions on serving memory footprint cannot exceed a certain input dimension and model size. This constraint raises a question on how to most effectively utilize a large-scale interaction corpus \cite{chang2023twin}. The most straightforward way is to truncate historical interactions. This simplification, however, comes at the cost of not using valuable information about user journeys and their rich history of interactions during model training \cite{Hua_2023}. Only focusing on the recent items provides little or no knowledge of the long tail of items in the catalog \cite{kitaev2020reformer}.

We hypothesize that the foundation model pretraining stage can benefit from receiving as much information as possible from interaction sequences. We then propose a method which involves curating input sequences for training samples using a hybrid method to include user item interactions from history along with most recent interactions. We provide an implementation approach using a sliding window sampling method to select portions of the interactions during each epoch. We also present quantitative and qualitative results that demonstrate the effectiveness of the method in learning long-term user preferences and in generating rich item representations.

\begin{figure}[h]
  \centering
  \includegraphics[scale=0.65]{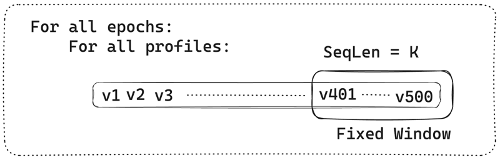}
  \caption{ Control training loop}
  \Description{Fixed window of length k is selected by truncating the large interaction sequence for all profiles in all epochs}
  \label{fig:control}
\end{figure}

\begin{table*}[h]
\small
\begin{tabular}{lcccccc}
\toprule
\textbf{Model}  & \textbf{Perplexity} & \textbf{MRR} & \textbf{mAP} & \textbf{Recall} \\
\midrule
Control     & -          & -         & -         & -         \\
All-Sliding  & -3.1\%     & -1.2\%    & +1.5\%    & +7.01\%   \\
Mixed-500    & -1.66\%    & +6.3\%    & \textbf{+18.29\%}  & +13.87\%  \\
Mixed-1000   & \textbf{-2.43\%}    & \textbf{+8.82\%}   & +14.04\%  & \textbf{+14.41\%}  \\
\bottomrule
\end{tabular}
\caption{Results}
\label{tab:results}
\end{table*}

\section{Related Work}
Foundation Models are now widely being adopted in RecSys \cite{geng2023recommendation}. Maximizing the interaction history of users is actively being explored to make learning more efficient \cite{zhu2024pose}. \cite{zhai2024actions} focuses on a setting with temporally repetitive user behaviors, enabling the use of artificial sparsification to reduce down long sequences. By contrast, we focus on data augmentation to improve the richness of representations under inference latency and model capacity constraints. Generally across large foundation models, there has been work on extending the context windows to a very large token size without affecting the model performance \cite{ding2024longrope} but is largely orthogonal to our goal of exploiting long sequence corpora without increasing model context windows.

\section{Approach}
\subsection{Model}
We choose a recommendation model released in practice as our baseline that follows the same auto-regressive prediction objective. This baseline model has an input sequence limit of 100 items to satisfy online serving latency constraints. This choice is purely illustrative, as the technique itself is general to any context window size.

\subsection{Dataset}
We use a large user interactions dataset of the order of 250M users and their interactions with the items in the content library. Example interaction sequences of users include video plays, video likes, add to watchlist, open video details page, etc. These interactions could span over long periods of time ranging from weeks to months. 

\subsection{Control Training Loop}
In the control training loop in baseline (Figure~\ref{fig:control}), we process each training example in a batch using a truncating sampler that selects the latest k=100 items in the user's interaction history. For users with less than K interactions, we take all of them. This method only focuses on the latest items the user interacted with.
\begin{figure}[h]
  \centering
  \includegraphics[scale=0.6]{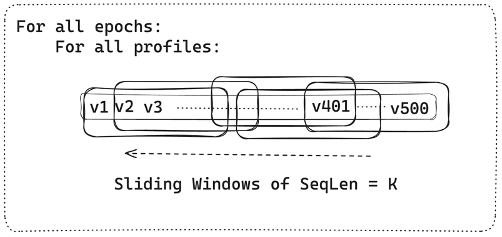}
  \caption{ Sliding window training loop}
  \Description{Sliding window of length k is selected by moving a sliding window over the large interaction sequence for all profiles in all epochs}
  \label{fig:treatment}
\end{figure}

\subsection{Sliding Window Training Loop}
In this loop, we process each training example in a batch using a sliding window sampler. It creates a sliding window of the same size of K=100 items and slides the window over different contiguous parts of the user history sequence. During each epoch, the sliding window function selects a different portion of the user-item history as input to the FM. Figure~\ref{fig:treatment} illustrates the sliding window training loop. This enables two key benefits - 
\begin{enumerate}
  \item more user interaction sequences are encoded, and the model learns more about users’ long-term interests, and
  \item the average number of items observed during training increases as compared to the fixed window approach where each epoch trains with the same user-item sequence.
\end{enumerate}

\subsection{Proposed Approach}
The primary task of RecSys FM is to recommend new items that the user is most interested in. Naturally, newer interactions are more immediately relevant to current behaviors than older ones. Older interactions are useful for the model to learn user’s interaction patterns and long-term interest. To balance these two objectives - recency and long term interests, we choose a mixed approach in the training loop - during entire training on $N$ epochs, we make $X$ epochs focus on the latest user item sequences using a fixed latest window and $N-X$ epochs focus on sliding window sampling that choose interactions from the entire history. After sweeping for an optimal $X$, we find that this mixed method works well in accomplishing both objectives.

\section{Results}

We compare the baseline fixed window RecSys FM model with $3$ treatment candidates -
\textit{All-Sliding} uses a sliding window training loop for all $N$ epochs without any focus on recent interactions. \textit{Mixed-500} uses a mixture of fixed recent \& sliding window epochs where sliding windows go as far back as $500$ item interaction events, and \textit{Mixed-1000} where we slide far back as $1000$ events.

We choose a combination of test sets containing recent interactions, random old item sequences \& unseen future held-out interactions after the training date. To assess the quality of item embedding  learnt, we use internal item similarity datasets. Our evaluation metrics are perplexity, Mean Reciprocal Rank \cite{Craswell2009} (MRR) for next item prediction and mAP \& recall for item embedding quality.

Our results in Table~\ref{tab:results} show effectiveness of the sliding window training approach over baseline on all evaluation metrics.

\section{Conclusions}
RecSys FMs that have tighter inference time limitations can maximize the utility of available user item interactions through a sliding window training approach for understanding long term user interests. Furthermore, this approach improves the overall quality of item representations learnt by the foundation model. 

Future work includes optimizing hyper parameters of this method, improving training time \& memory efficiency while sliding through smarter parallelization and lastly, exploring of new sliding window functions that learn different aspects of user interests.

\section{Author Bios}
\textbf{Swanand Joshi} is an applied machine learning researcher at Netflix, focusing on personalization algorithms and member satisfaction.

\textbf{Yesu Feng} is a research scientist at Netflix working on foundation models and interactive recommendations.

\textbf{Ko-Jen Hsiao} is a staff research scientist/engineer at Netflix, with 7 years of experience working on Netflix's core recommendation systems. His current focus is on developing foundational models for various algorithms that drive Netflix's personalized experiences. He earned his PhD from the University of Michigan, where he specialized in combining disparate information for machine learning applications. 

\textbf{Zhe Zhang}, Ph.D. in Computer Science, is a senior research scientist at Netflix. Previously, he was a research scientist at Meta AI and IBM, working on NLP, dialog systems and recommendations.

\textbf{Sudarshan Lamkhede} is the Engineering Manager of the Machine Learning - Foundation Models, Search, and Recommendations at Netflix Research. Prior to Netflix, he led research engineering for Web Search algorithms at Yahoo! Research. He co-organizes the San Francisco Bay Area Machine Learning symposium (BayLearn).






\bibliographystyle{ACM-Reference-Format}
\bibliography{fm-paper-refs}


\begin{thebibliography}{11}


\ifx \showCODEN    \undefined \def \showCODEN     #1{\unskip}     \fi
\ifx \showDOI      \undefined \def \showDOI       #1{#1}\fi
\ifx \showISBNx    \undefined \def \showISBNx     #1{\unskip}     \fi
\ifx \showISBNxiii \undefined \def \showISBNxiii  #1{\unskip}     \fi
\ifx \showISSN     \undefined \def \showISSN      #1{\unskip}     \fi
\ifx \showLCCN     \undefined \def \showLCCN      #1{\unskip}     \fi
\ifx \shownote     \undefined \def \shownote      #1{#1}          \fi
\ifx \showarticletitle \undefined \def \showarticletitle #1{#1}   \fi
\ifx \showURL      \undefined \def \showURL       {\relax}        \fi
\providecommand\bibfield[2]{#2}
\providecommand\bibinfo[2]{#2}
\providecommand\natexlab[1]{#1}
\providecommand\showeprint[2][]{arXiv:#2}

\bibitem[Chang et~al\mbox{.}(2023)]%
        {chang2023twin}
\bibfield{author}{\bibinfo{person}{Jianxin Chang}, \bibinfo{person}{Chenbin Zhang}, \bibinfo{person}{Zhiyi Fu}, \bibinfo{person}{Xiaoxue Zang}, \bibinfo{person}{Lin Guan}, \bibinfo{person}{Jing Lu}, \bibinfo{person}{Yiqun Hui}, \bibinfo{person}{Dewei Leng}, \bibinfo{person}{Yanan Niu}, \bibinfo{person}{Yang Song}, {and} \bibinfo{person}{Kun Gai}.} \bibinfo{year}{2023}\natexlab{}.
\newblock \bibinfo{title}{TWIN: TWo-stage Interest Network for Lifelong User Behavior Modeling in CTR Prediction at Kuaishou}.
\newblock
\newblock
\showeprint[arxiv]{2302.02352}~[cs.IR]


\bibitem[Craswell(2009)]%
        {Craswell2009}
\bibfield{author}{\bibinfo{person}{Nick Craswell}.} \bibinfo{year}{2009}\natexlab{}.
\newblock \bibinfo{booktitle}{\emph{Mean Reciprocal Rank}}.
\newblock \bibinfo{publisher}{Springer US}, \bibinfo{address}{Boston, MA}, \bibinfo{pages}{1703--1703}.
\newblock
\showISBNx{978-0-387-39940-9}
\urldef\tempurl%
\url{https://doi.org/10.1007/978-0-387-39940-9_488}
\showDOI{\tempurl}


\bibitem[Ding et~al\mbox{.}(2024)]%
        {ding2024longrope}
\bibfield{author}{\bibinfo{person}{Yiran Ding}, \bibinfo{person}{Li~Lyna Zhang}, \bibinfo{person}{Chengruidong Zhang}, \bibinfo{person}{Yuanyuan Xu}, \bibinfo{person}{Ning Shang}, \bibinfo{person}{Jiahang Xu}, \bibinfo{person}{Fan Yang}, {and} \bibinfo{person}{Mao Yang}.} \bibinfo{year}{2024}\natexlab{}.
\newblock \bibinfo{title}{LongRoPE: Extending LLM Context Window Beyond 2 Million Tokens}.
\newblock
\newblock
\showeprint[arxiv]{2402.13753}~[cs.CL]


\bibitem[Geng et~al\mbox{.}(2023)]%
        {geng2023recommendation}
\bibfield{author}{\bibinfo{person}{Shijie Geng}, \bibinfo{person}{Shuchang Liu}, \bibinfo{person}{Zuohui Fu}, \bibinfo{person}{Yingqiang Ge}, {and} \bibinfo{person}{Yongfeng Zhang}.} \bibinfo{year}{2023}\natexlab{}.
\newblock \bibinfo{title}{Recommendation as Language Processing (RLP): A Unified Pretrain, Personalized Prompt `I\&' Predict Paradigm (P5)}.
\newblock
\newblock
\showeprint[arxiv]{2203.13366}~[cs.IR]


\bibitem[Hua et~al\mbox{.}(2023)]%
        {Hua_2023}
\bibfield{author}{\bibinfo{person}{Wenyue Hua}, \bibinfo{person}{Shuyuan Xu}, \bibinfo{person}{Yingqiang Ge}, {and} \bibinfo{person}{Yongfeng Zhang}.} \bibinfo{year}{2023}\natexlab{}.
\newblock \showarticletitle{How to Index Item IDs for Recommendation Foundation Models}. In \bibinfo{booktitle}{\emph{Proceedings of the Annual International ACM SIGIR Conference on Research and Development in Information Retrieval in the Asia Pacific Region}} \emph{(\bibinfo{series}{SIGIR-AP ’23})}. \bibinfo{publisher}{ACM}.
\newblock
\urldef\tempurl%
\url{https://doi.org/10.1145/3624918.3625339}
\showDOI{\tempurl}


\bibitem[Huang et~al\mbox{.}(2024)]%
        {huang2024foundation}
\bibfield{author}{\bibinfo{person}{Chengkai Huang}, \bibinfo{person}{Tong Yu}, \bibinfo{person}{Kaige Xie}, \bibinfo{person}{Shuai Zhang}, \bibinfo{person}{Lina Yao}, {and} \bibinfo{person}{Julian McAuley}.} \bibinfo{year}{2024}\natexlab{}.
\newblock \bibinfo{title}{Foundation Models for Recommender Systems: A Survey and New Perspectives}.
\newblock
\newblock
\showeprint[arxiv]{2402.11143}~[cs.IR]


\bibitem[Kitaev et~al\mbox{.}(2020)]%
        {kitaev2020reformer}
\bibfield{author}{\bibinfo{person}{Nikita Kitaev}, \bibinfo{person}{Łukasz Kaiser}, {and} \bibinfo{person}{Anselm Levskaya}.} \bibinfo{year}{2020}\natexlab{}.
\newblock \bibinfo{title}{Reformer: The Efficient Transformer}.
\newblock
\newblock
\showeprint[arxiv]{2001.04451}~[cs.LG]


\bibitem[Sun et~al\mbox{.}(2019)]%
        {DBLP:journals/corr/abs-1904-06690}
\bibfield{author}{\bibinfo{person}{Fei Sun}, \bibinfo{person}{Jun Liu}, \bibinfo{person}{Jian Wu}, \bibinfo{person}{Changhua Pei}, \bibinfo{person}{Xiao Lin}, \bibinfo{person}{Wenwu Ou}, {and} \bibinfo{person}{Peng Jiang}.} \bibinfo{year}{2019}\natexlab{}.
\newblock \showarticletitle{BERT4Rec: Sequential Recommendation with Bidirectional Encoder Representations from Transformer}.
\newblock \bibinfo{journal}{\emph{CoRR}}  \bibinfo{volume}{abs/1904.06690} (\bibinfo{year}{2019}).
\newblock
\showeprint[arXiv]{1904.06690}
\urldef\tempurl%
\url{http://arxiv.org/abs/1904.06690}
\showURL{%
\tempurl}


\bibitem[Xia et~al\mbox{.}(2023)]%
        {xia2023transact}
\bibfield{author}{\bibinfo{person}{Xue Xia}, \bibinfo{person}{Pong Eksombatchai}, \bibinfo{person}{Nikil Pancha}, \bibinfo{person}{Dhruvil~Deven Badani}, \bibinfo{person}{Po-Wei Wang}, \bibinfo{person}{Neng Gu}, \bibinfo{person}{Saurabh~Vishwas Joshi}, \bibinfo{person}{Nazanin Farahpour}, \bibinfo{person}{Zhiyuan Zhang}, {and} \bibinfo{person}{Andrew Zhai}.} \bibinfo{year}{2023}\natexlab{}.
\newblock \bibinfo{title}{TransAct: Transformer-based Realtime User Action Model for Recommendation at Pinterest}.
\newblock
\newblock
\showeprint[arxiv]{2306.00248}~[cs.IR]


\bibitem[Zhai et~al\mbox{.}(2024)]%
        {zhai2024actions}
\bibfield{author}{\bibinfo{person}{Jiaqi Zhai}, \bibinfo{person}{Lucy Liao}, \bibinfo{person}{Xing Liu}, \bibinfo{person}{Yueming Wang}, \bibinfo{person}{Rui Li}, \bibinfo{person}{Xuan Cao}, \bibinfo{person}{Leon Gao}, \bibinfo{person}{Zhaojie Gong}, \bibinfo{person}{Fangda Gu}, \bibinfo{person}{Michael He}, \bibinfo{person}{Yinghai Lu}, {and} \bibinfo{person}{Yu Shi}.} \bibinfo{year}{2024}\natexlab{}.
\newblock \bibinfo{title}{Actions Speak Louder than Words: Trillion-Parameter Sequential Transducers for Generative Recommendations}.
\newblock
\newblock
\showeprint[arxiv]{2402.17152}~[cs.LG]


\bibitem[Zhu et~al\mbox{.}(2024)]%
        {zhu2024pose}
\bibfield{author}{\bibinfo{person}{Dawei Zhu}, \bibinfo{person}{Nan Yang}, \bibinfo{person}{Liang Wang}, \bibinfo{person}{Yifan Song}, \bibinfo{person}{Wenhao Wu}, \bibinfo{person}{Furu Wei}, {and} \bibinfo{person}{Sujian Li}.} \bibinfo{year}{2024}\natexlab{}.
\newblock \bibinfo{title}{PoSE: Efficient Context Window Extension of LLMs via Positional Skip-wise Training}.
\newblock
\newblock
\showeprint[arxiv]{2309.10400}~[cs.CL]


\end{thebibliography}

\end{document}